\documentclass[preprint, authoryear,12pt]{elsarticle}

\usepackage{amssymb}
\usepackage{amsmath}
\usepackage[amsthm]{ntheorem}
\usepackage[utf8]{inputenc}

 \usepackage{natbib}	

\usepackage{booktabs}
\usepackage{multirow}
\usepackage{tabularx}

\usepackage{dsfont}

\usepackage{lipsum}

\usepackage{graphicx}

\usepackage{placeins}

\usepackage{setspace}

\usepackage[textsize=tiny]{todonotes}

\newcommand{\var}{\operatorname{var}}
\newcommand{\cov}{\operatorname{cov}}

\newtheorem{Exa}{Example}

\journal{Computational Statistics \& Data Analysis}

\begin{document}

\begin{frontmatter}

\title{Measuring Association between Random Vectors}
\author[lehrstuhl]{Oliver Grothe\corref{cor1}}
\ead{grothe@statistik.uni-koeln.de}
\author[lehrstuhl]{Friedrich Schmid}
\ead{schmid@wiso.uni-koeln.de}
\author[gk]{Julius Schnieders}
\ead{schnieders@wiso.uni-koeln.de}
\author[belgien]{Johan Segers}
\ead{johan.segers@uclouvain.be}

\cortext[cor1]{Corresponding author}

\address[lehrstuhl]{University of Cologne, Department of Economic and Social Statistics, Albertus-Magnus-Platz, 50923 Cologne, Germany}
\address[gk]{University of Cologne, Graduate School of Risk Management, Meister-Ekkehart-Strasse 11, 50923 Cologne, Germany}
\address[belgien]{Universit\'e catholique de Louvain, Institut de statistique, biostatistique et sciences actuarielles, Voie du Roman Pays 20,
B-1348 Louvain-la-Neuve, Belgium}

\begin{abstract}
\noindent
This paper suggests five measures of association between two random vectors $\textbf{X} = (X_1,\ldots,X_p)$ and
$\textbf{Y} = (Y_1,\ldots,Y_q)$. They are copula based and therefore invariant with respect to the marginal
distributions of the components $X_i$ and $Y_j$. The measures capture positive as well as negative association of $\textbf{X}$ and $\textbf{Y}$.
In case $p=q=1$ they reduce to Spearman's rho. Various properties of these new measures are investigated. Nonparametric estimators, based on ranks, for the measures are derived and their small sample behavior is investigated by simulation. The measures are applied to characterise strength and direction of association of bond and stock indices of five countries over time.
\end{abstract}

\end{frontmatter}

\newpage

\section{Introduction}

Association of two random variables $X$ and $Y$ has been thoroughly investigated in the statistical literature but much less work concerns
association of two random vectors $\mathbf{X}=\left( X_{1},\ldots,X_{p}\right)$ and $\mathbf{Y}=\left( Y_{1},\ldots,Y_{q}\right) $. In the classical
statistical framework of multivariate normality and linear correlation the theory of canonical correlation (see \cite{Hotelling1936}) and the socalled RV
coefficient (see \cite{Escoufier1973} and \cite{Robert1976}) are well known and often applied, in particular in the natural sciences.

The focus of our application, however, is financial data which is
notoriously non-normal. It has further been pointed out (see \cite{Embrechts2002}) that linear correlation might be inappropriate to measure the
strength of association of financial data. Finally the above mentioned
measures are not capable of distinguishing between positive and negative
association which is a must in dealing with financial data.

The recently introduced distance correlation (see \cite{Szekely2007} and \cite{Szekely2009}) suffers from similar weaknesses. Further it depends on the marginal distributions of $\mathbf{X}$ and $\mathbf{Y}$ and requires moment restrictions.
This should be considered as a disadvantage in application to financial data (see again \cite{Embrechts2002} and \cite{Remillard2009b}).

The increasing use of copulas in the analysis and modeling of financial data suggests the development of copula based measures of association.
Concerning association \textit{within} one vector $\mathbf{X}=\left( X_{1},\ldots,X_{p}\right),$ there exist various such measures (see \cite{Schmid2009} for a recent survey). In this paper, we propose copula based measures of association \textit{between}
two vectors $\mathbf{X}=\left( X_{1},\ldots,X_{p}\right) $ and $\mathbf{Y}%
=\left( Y_{1},\ldots,Y_{q}\right) .$ They are invariant with respect to
marginal distributions as association is solely determined by the joint
copula of $\mathbf{X}$ and $\mathbf{Y}$. Therefore distributional
assumptions (beside continuity) as well as moment restrictions are not necessary.
The measures are further capable of measuring negative association.

The measures are defined in such a way that they reduce to Spearman's rank correlation coefficient for $p=q=1.$ Instead of Spearman's coefficient, other measures, such as Kendall's coefficient, could have been used. The details of this approach are very similar and are omitted.

Note
that it is not the aim of this paper to derive tests for independence of $%
\mathbf{X}$ and $\mathbf{Y}$ (see \cite{Beran_etal_2007}, \cite{Kojadinovic2009} and \cite{Quessy2010} for recent contributions). On the contrary the focus is measurement
of type and strength of association in the case where $\mathbf{X}$ and $\mathbf{Y}$ are dependent.

The structure of the paper is as follows: In section \ref{sec.notdef}, we describe the notation and terminology used within the paper. In section \ref{sec.popver}, we state population versions of the five copula based measures of association and derive some of their  properties. For convenience, some calculations are collected in the appendix.
Estimators for the measures are introduced in section \ref{sec.empest}. Their finite sample properties are analysed in a simulation study. Matlab code for the estimators is available on request. Section \ref{sec.empex} contains an empirical example with financial data, illustrating the usefulness of the new measures. Section \ref{sec.concl} concludes.

\section{Notation and definitions}
\label{sec.notdef}
Let $\mathbf{X}=\left( X_{1},\ldots,X_{p}\right) $ and $\mathbf{Y}=\left(Y_{1},\ldots,Y_{q}\right) $ be random vectors of dimensions $p$ and $q$, respectively, defined on the same probability space. Throughout the paper we assume that the marginal distribution functions $F_{X_{i}}$ for $i=1,\ldots,p$ and $F_{Y_{j}}$ for $j=1,\ldots,q$ are continuous functions. Therefore, according to the theorem of \cite{Sklar1959} there exists a unique copula
$C:\left[ 0,1\right] ^{p+q}\longrightarrow \left[ 0,1\right]$
with
\begin{multline*}
P\left( X_{1}\leq x_{1},\ldots,X_{p}\leq x_{p},Y_{1}\leq y_{1},\ldots,Y_{q} \leq y_{q}\right) \\
 =C\left( F_{X_{1}}\left( x_{1}\right) ,\ldots,F_{X_{p}}\left( x_{p}\right)
,F_{Y_{1}}\left( y_{1}\right) ,\ldots,F_{Y_{q}}\left( y_{q}\right) \right)
\end{multline*}
for $\left( x_{1},\ldots,x_{p}\right) \in \mathbb{R}^p$ and $\left( y_{1},\ldots,y_{q}\right) \in \mathbb{R}^q$.
Extensive portrayals of copulas are given in \citet{Nelsen2006}, \citet{Joe1997} and \citet{Cherubinietal2004}. Let
\begin{equation*}
\mathbf{U}=\left( U_{1},\ldots,U_{p}\right) =\left( F_{X_{1}}\left(
X_{1}\right) ,\ldots,F_{X_{p}}\left( X_{p}\right) \right)
\end{equation*}%
and%
\begin{equation*}
\mathbf{V}=\left( V_{1},\ldots,V_{q}\right) =\left( F_{Y_{1}}\left(
Y_{1}\right) ,\ldots,F_{Y_{q}}\left( Y_{q}\right) \right) .
\end{equation*}
The copula $C$ is the joint distribution function of $\left( \mathbf{U},\mathbf{V}\right)$. The marginal copulas of $\mathbf{X}$ and $\mathbf{Y}$ are given by $A\left(\mathbf{u}\right) =C\left( \mathbf{u},\mathbf{1}_{q}\right) $ \ and $B\left(\mathbf{v}\right) =C\left(\mathbf{1}_{p},\mathbf{v}\right)$ for $\mathbf{u}=\left( u_{1},\ldots,u_{p}\right) \in \left[ 0,1\right] ^{p}$ and $\mathbf{v}=\left( v_{1},\ldots,v_{q}\right) \in \left[ 0,1\right] ^{q}$. Here, $\mathbf{1}_{p}$ and $\mathbf{1}_{q}$ denote vectors of ones of length $p$
and $q$, respectively. By construction, $A$ and $B$ are the marginal distribution
functions of $\mathbf{U}$ and $\mathbf{V}$, respectively.

\section{Population versions of measures of association}
\label{sec.popver}
This section introduces population versions of five copula based measures of association between random vectors $\mathbf{X}$ and $\mathbf{Y}$.
\subsection{Mean of pairwise association}

The simplest measure of association between $\mathbf{X}$ and $\mathbf{Y}$
is the mean of all bivariate associations of $X_{i}$ and
$Y_{j}$ for \ $i=1,\ldots,p$ \ and \ $j=1,\ldots,q$, \ i.e.,
\begin{equation*}
\overline{\rho }_{\mathbf{X},\mathbf{Y}}=\frac{1}{pq} \sum_{i=1}^{p} \sum_{j=1}^{q} \rho _{ij}
\end{equation*}%
where $\rho _{ij}$ is Spearman's rho of $X_{i}$ and $Y_{j}$. Note that $\rho
_{ij}$ is copula based because of%
\begin{equation*}
\rho _{ij}= 12 \int_{\left[0,1\right]} \int_{\left[0,1\right]} C_{ij}\left(
u_{i},v_{j}\right) \, du_{i} \, dv_{j}-3
\end{equation*}
and $C_{ij}$ is the marginal copula of $X_{i}$ and $Y_{j}$ (see \cite{Nelsen2006}).

The properties of $\overline{\rho }_{\mathbf{X},\mathbf{Y}}$ are easy to
derive and follow directly from its definition.

\begin{enumerate}
\item We have $-1\leq \overline{\rho }_{\mathbf{X},\mathbf{Y}}\leq 1$
and the measure is invariant with respect to permutations within $\mathbf{X}$ and $\mathbf{Y}$.
\item $\overline{\rho }_{\mathbf{X},\mathbf{Y}}=-1$ if and only if $\rho
_{ij}=-1$ for every combination of $i$ and $j$, i.e., $X_{i}$ and $Y_{j}$ are countermonotonic for all $i$ and $j$. This implies that
$$\mathbf{X}=\left( h_{1}\left( X\right) ,\ldots,h_{p}\left( X\right)
\right)\,\,\,\text{and}\,\,\,\mathbf{Y}=\left( g_{1}\left( X\right) ,\ldots,g_{q}\left( X\right)
\right)$$%
for a random variable $X$ and strictly increasing functions $h_{i},i=1,\ldots,p$
and strictly decreasing functions $g_{j},j=1,\ldots,q$, and it follows that
\begin{equation*}
C\left( \mathbf{u},\mathbf{v}\right) =\max \left\{ 0,\left( \min \left\{
u_{1},\ldots,u_{p}\right\} +\min \left\{ v_{1},\ldots,v_{q}\right\} -1\right)
\right\} .
\end{equation*}
\item $\overline{\rho }_{\mathbf{X},\mathbf{Y}}=1$ if and only if $\rho
_{ij}=1$ for every combination $i=1,\ldots,p$ \ and \ $j=1,\ldots,q$. Here,
$X_{i}$ and $X_{j}$ are comonotonic for all $i$ and $j$. This implies that%
$$\mathbf{X}=\left( h_{1}\left( X\right) ,\ldots,h_{p}\left( X\right)
\right)\,\,\,\text{and}\,\,\,\mathbf{Y}=\left( g_{1}\left( X\right) ,\ldots,g_{q}\left( X\right)
\right)$$%
for strictly increasing functions $h_{i},i=1,\ldots,p$ and $g_{j},j=1,\ldots,q$
for a random variable $X$. The copula $C$ is then
\begin{equation*}
C\left( \mathbf{u,v}\right) =\min \left\{
u_{1},\ldots,u_{p},v_{1},\ldots,v_{q}\right\} .
\end{equation*}

For given and fixed marginal copulas $A$ and $B$ it is in
general not possible to find a copula $C$ with $A(\mathbf{u})=C\left(
\mathbf{u},\mathbf{1}_q\right) $ \ and \ $B(\mathbf{v})=C\left( \mathbf{1}_p,v\right)$
which entails $\overline{\rho }=-1$ \ or \ $\overline{\rho }=+1.$
The latter cases are only possible in the special cases \ $A(\mathbf{u}%
)=\min \left\{ u_{1},\ldots,u_{p}\right\} $ \ and \ $B(\mathbf{v})=\min \left\{
v_{1},\ldots,v_{q}\right\} .$

\item If $X_{i}$ and $Y_{j}$ are independent for all combinations $i=1,\ldots,p$
\ and \ $j=1,\ldots,q$ then $\overline{\rho }_{\mathbf{X},\mathbf{Y}}=0$. The
converse is not true as there may be some $\rho _{ij}$ different from zero, but $\overline{\rho }=0$:
\begin{Exa}
Consider two 2-dimensional random vectors $\textbf{X} = (X_1,X_2)$ and $\textbf{Y} = (Y_1, Y_2)$ with the following
matrix of Spearman's rank correlations:
\begin{equation*}
\Theta =
\left(
\begin{array}{cccc}
	1 & a	&	b	& b 	\\
	a	&	1	& -b & -b \\
	b & -b & 1 & a \\
	b & -b & a & 1
\end{array}
\right)
\end{equation*}
Choosing, for example,  $a=0.6$ and $b=0.4$, the matrix $\Theta$ is positive semi-definite and thus a correlation matrix. In this case, the random vectors $\textbf{X}$ and $\textbf{Y}$ are clearly dependent, whereas the mean of the pairwise associations is equal to
\begin{equation*}
\overline{\rho} = \frac{1}{4} \left( \rho_{13} + \rho_{14} + \rho_{23} + \rho_{24}\right) = 0.
\end{equation*}
\end{Exa}


\item Association as measured by $\overline{\rho}$ can be decomposed into two parts that describe the association within and between the two random vectors.
Let $\left( \mathbf{X},\mathbf{Y}\right) =\mathbf{Z}=\left(Z_{1},\ldots,Z_{p+q}\right) $ and let $\overline{\rho }_{\mathbf{Z}}$ denote
total association \textit{within} $\mathbf{Z}$ defined by
\begin{equation*}
\overline{\rho }_{\mathbf{Z}} = \binom{p+q}{2}^{-1} \sum_{\substack{l<r\\l,r\in\left\{ 1,\ldots,p+q\right\}}} \rho _{lr}
\end{equation*}
where $\rho _{lr}$ denotes Spearman's rho of $Z_{l}$ and $Z_{r}$. Then%

\begin{eqnarray*}
\overline{\rho }_{\mathbf{Z}} & = & \binom{p+q}{2}^{-1}\left\{ \sum_{l=1}^{p} \sum_{r=p+1}^{q} \rho_{lr}
+  \sum_{\substack{l<r\\l,r\in\left\{ 1,\ldots,p\right\}}} \rho_{lr}
+  \sum_{\substack{l<r\\l,r\in\left\{ p+1,\ldots,p+q\right\}}} \rho_{lr} \right\} \\ \displaystyle
& = & \underbrace{\frac{pq}{\binom{p+q}{2}} \overline{\rho }_{\mathbf{X},\mathbf{Y}}}_{\text{between}}+\underbrace{\frac{\binom{p}{2}}{\binom{p+q}{2}}\overline{\rho }_{\mathbf{X%
}}+\frac{\binom{q}{2}}{\binom{p+q}{2}}\overline{\rho }_{\mathbf{Y}}.}_{\text{within}}
\end{eqnarray*}
Note that%
\begin{equation*}
\frac{pq}{\binom{p+q}{2}}+\frac{\binom{p}{2}}{\binom{p+q}{2}}+\frac{%
\binom{q}{2}}{\binom{p+q}{2}}=1.
\end{equation*}%
\end{enumerate}
This property might be useful since decomposition of $\overline{\rho }_{\mathbf{Z}}$ into a between and within
part can be interesting in analysing financial data.

\subsection{Pearson correlation based measures of association}
\label{sec.pearsmeas}
The vectors $\mathbf{X}$ and $\mathbf{Y}$ are independent if and only if  $$%
C\left( \mathbf{u},\mathbf{v}\right) =A\left( \mathbf{u}\right) B\left(
\mathbf{v}\right) $$ for $\mathbf{u}\in \left[ 0,1\right] ^{p}$  and $%
\mathbf{v}\in \left[ 0,1\right] ^{q}$.
Equivalently, $\mathbf{X}$ and $\mathbf{Y}$ are independent if and only if
\begin{equation*}
\overline{C}\left( \mathbf{u},\mathbf{v}\right) =P\left( \mathbf{U}\geqslant
\mathbf{u},\mathbf{V}\geqslant \mathbf{v}\right) =P\left( \mathbf{U}%
\geqslant \mathbf{u}\right) P\left( \mathbf{V}\geqslant \mathbf{v}\right) =%
\overline{A}\left( \mathbf{u}\right) \overline{B}\left( \mathbf{v}\right)
\end{equation*}%
for \ $\mathbf{u}\in \left[ 0,1\right] ^{p}$  and  $\mathbf{v}\in \left[
0,1\right] ^{q}$. Therefore, we may measure the distance of $C$ to the independence case of $\textbf{X}$ and $\textbf{Y}$ by
\begin{equation*}
\int_{\left[0,1\right]^p} \int_{\left[0,1\right]^q} \left(C\left(
\mathbf{u},\mathbf{v}\right) -A\left( \mathbf{u}\right) B\left( \mathbf{v}%
\right)\right) \, d\mathbf{v} \, d\mathbf{u}.
\end{equation*}%
This expression is equal to
\begin{equation*}
\cov\left( \pi _{p}\left( \mathbf{U}\right) ,\pi _{q}\left( \mathbf{V}\right)
\right)
\end{equation*}%
where
\begin{equation*}
\pi _{p}\left( \mathbf{u}\right) =\underset{i=1}{\overset{p}{\prod }}\left(
1-u_{i}\right) \text{ \ and \ }\pi _{q}\left( \mathbf{v}\right) =\underset{%
j=1}{\overset{q}{\prod }}\left( 1-v_{j}\right)
\end{equation*}%
for $\mathbf{u}\in \left[ 0,1\right] ^{p}$ \ and \ $\mathbf{v}\in \left[ 0,1%
\right] ^{q}$ (see Appendix I for a proof).

The covariance is bounded by the product of the respective standard deviations. Thus, a measure of association between $\mathbf{X%
}$ and $\mathbf{Y}$ which is bounded by $-1$ and $1$ is%
\begin{equation*}
\rho _{1}\left( \mathbf{X},\mathbf{Y}\right) =\rho _{1}\left( C\right) =%
\frac{\cov\left( \pi _{p}\left( \mathbf{U}\right) ,\pi _{q}\left( \mathbf{V}%
\right) \right) }{\sqrt{\var\left( \pi _{p}\left( \mathbf{U}\right) \right) }%
\sqrt{\var\left( \pi _{q}\left( \mathbf{V}\right) \right) }}.
\end{equation*}%
The variances may be expressed by %
\begin{equation*}
\var\left( \pi _{p}\left( \mathbf{U}\right) \right) =
\int_{\left[0,1\right]^p} \int_{\left[0,1\right]^p} \left( A\left( \mathbf{u}\wedge \mathbf{u}%
^{\prime }\right) -A\left( \mathbf{u}\right) A\left( \mathbf{u}^{\prime }\right)
\right) d\mathbf{u}^{\prime }d\mathbf{u},
\end{equation*}%
analogously for $\var\left( \pi _{q}\left( \mathbf{V}\right) \right)$ (see Appendix I for a derivation). By $\mathbf{u} \wedge \mathbf{u}' := (\min\{u_1,u_1'\}, \dots, \min\{u_p,u_p'\})$ we denote the component wise minimum of $\mathbf{u}$ and $\mathbf{u}'$.

The measure $\rho_1$ is based on the covariance of $\mathbf{U}$ and $\mathbf{V}$ transformed by the functions $\pi_{p}$ and $\pi_{q}.$ An alternative approach is to transform $\mathbf{U}$ and $\mathbf{V}$ by their respective distribution functions $A$ and $B$ (see \cite{Nelsen2003}) and consider
the measure of association defined by%
$$
\rho _{2}\left( \mathbf{X},\mathbf{Y}\right) =\rho _{2}(C)=\frac{\cov\left(
A\left( \mathbf{U}\right) ,B\left( \mathbf{V}\right) \right) }{\sqrt{%
var\left( A\left( \mathbf{U}\right) \right) }\sqrt{var\left( B\left( \mathbf{%
V}\right) \right) }}.$$
It is shown in Appendix I that
\begin{equation*}
\cov\left(
A\left( \mathbf{U}\right) ,B\left( \mathbf{V}\right) \right)=\int_{\left[0,1\right]^p} \int_{\left[0,1\right]^q} \left(
\overline{C}\left( \mathbf{u},\mathbf{v}\right) -\overline{A}\left( \mathbf{u%
}\right) \overline{B}\left( \mathbf{v}\right) \right) \, dB\left( \mathbf{v}%
\right) \, dA\left( \mathbf{u}\right)
\end{equation*}%
and
\begin{equation*}
\var\left( A\left( \mathbf{U}\right) \right) =\int_{\left[0,1\right]^p} \int_{\left[0,1\right]^q}
\left( \overline{A}\left( \mathbf{u}\vee
\mathbf{u}^{\prime }\right) -\overline{A}\left( \mathbf{u}\right) \overline{A%
}\left( \mathbf{u}^{\prime }\right) \right) \, dA\left( \mathbf{u}\right) \, dA\left( \mathbf{u}^{\prime }\right)
\end{equation*}%
with an analogous expression for $\var\left( B\left( \mathbf{V}\right) \right).$ By $\mathbf{u} \vee \mathbf{u}'$ := $(\max\{u_1,u_1'\}, \dots,$ $ \max\{u_p,u_p'\})$ we denote component wise maximum of $\mathbf{u}$ and $\mathbf{u}'$.

The properties of $\rho_1$ and $\rho_2$ are as follows:
\begin{enumerate}
\item We have $-1\leq {\rho }_{1}\leq 1$ and $-1\leq {\rho }_{2}\leq 1$
and the measures are invariant with respect to permutations within $\mathbf{X}$ and $\mathbf{Y}$.
\item If the vectors $\mathbf{X}$ and $\mathbf{Y}$ are independent, $\rho_1=0$ and $\rho_2=0.$ Again, the converse is not true.
\item Measures $\rho_1$ and $\rho_2$ are in general not capable to achieve $+1$ or $-1$ for given and fixed arbitrary copulas $A$ and $B$ of $\mathbf{X}$ and $\mathbf{Y}.$
To calculate the maximal and minimal values
of, for example, $\rho_1 (\mathbf{X},\mathbf{Y})$ for given copulas $A$ and $B$  we have to maximise and minimise $\rho_1 (\mathbf{X},\mathbf{Y})$ over all random vectors $(\mathbf{X},\mathbf{Y}),$ where the copula
of $\mathbf{X}$ is $A$ and the one of $\mathbf{Y}$ is $B$.
For fixed $A$ and $B$, the only term in the definition of $\rho_1 (\mathbf{X},\mathbf{Y})$ that can vary is
\begin{eqnarray*}
\int_{[0,1]^{p+q}} C(\textbf{u},\textbf{v}) \,d\textbf{u} \,d\textbf{v} & = & E \left[ \pi_p (\textbf{U}) \pi_q (\textbf{V}) \right] \\
& = & \int_{[0,1]^2} P \left[ \pi_p (\textbf{U}) > s,\pi_q (\textbf{V}) > t  \right] \, ds \, dt.
\end{eqnarray*}
The joint survival function of $\pi_p (\textbf{U})$ and $\pi_q (\textbf{V})$ will be maximal (minimal) if they are comonotone (countermonotone):
\begin{multline*}
\int_{[0,1]^{p+q}} C(\textbf{u},\textbf{v}) \,d\textbf{u} \,d\textbf{v}  \\
\ge \int_{[0,1]^2} \max \left( P \left[ \pi_p (\textbf{U}) > s \right] + P \left[ \pi_p (\textbf{V}) > t \right] -1,0\right) \, ds \, dt,
\end{multline*}
\begin{multline*}
\int_{[0,1]^{p+q}} C(\textbf{u},\textbf{v}) \,d\textbf{u} \,d\textbf{v}  \\
\le \int_{[0,1]^2} \min \left( P \left[ \pi_p (\textbf{U}) > s \right], P \left[ \pi_p (\textbf{V}) > t \right] \right) \, ds \, dt.
\end{multline*}
\begin{Exa}
Consider two 2-dimensional random vectors $\mathbf{X} = ( X_1,X_2)$ and $\mathbf{Y} = (Y_1,Y_2)$ with copulas
$C_{\mathbf{X}} (u_1, u_2) = \Pi(u_1, u_2)$ and $C_{\mathbf{Y}} (v_1, v_2) = W (v_1, v_2)$, where $\Pi(u_1, u_2) := u_1 u_2$ is
the independence copula and $W (v_1, v_2) := \max \left\{ v_1 + v_2 -1 ,0 \right\}$ the countermonotone copula. In this case

\begin{eqnarray*}
	P \left[ \pi_2 (\textbf{U}) > s \right] & = & P \left[ (1-U_1)(1-U_2) > s \right] \nonumber \\
	& = & \int_{0}^{1-s} P \left( U_2 < \left. \frac{1-s-u_1}{1-u_1} \right| U_1 = u_1 \right) \, du_1 \nonumber \\
	& = & \int_{0}^{1-s} \dot{\Pi}_1 \left(u_1, \frac{1-s-u_1}{1-u_1}\right) \, du_1 \nonumber \\
	& = & \int_{0}^{1-s} \frac{1-s-u_1}{1-u_1} \, du_1 \nonumber \\
	& = & 1 - s + s \log(s),
\end{eqnarray*}
and
\begin{eqnarray*}
 P \left[ \pi_2 (\textbf{V}) > t \right] & = & P \left[ (1-V_1)(1-V_2) > t \right] \nonumber \\
 & = & P \left[ (1-V_1) V_1 > t \right] \\
 & = &   2 \sqrt{ \max \left\{ \frac{1}{4} - t,0\right\}}.
\end{eqnarray*}																
The upper bound is given by
\begin{align*}
 \int_{[0,1]^4} C(\mathbf{u},\mathbf{v}) \,d\textbf{u} \,d\textbf{v}  \nonumber \\
 & \le \int_{[0,1]^2} \min \left( P \left[ \pi_2(\textbf{U} > s \right], P \left[ \pi_2 (\textbf{V}) > t \right] \right) \, ds \,dt \nonumber \\
 &  =  \int_{[0,1]^2} \min \left( 1 -s + s \log(s), 2 \sqrt{ \max \left\{ \frac{1}{4} - t,0\right\}}  \right) \, ds \,dt \nonumber \\
 &  = \frac{767}{13824} \approx 0.0555
\end{align*}
Thus, the maximal association between $\textbf{X}$ and $\textbf{Y}$ in terms of $\rho_1$ is
\begin{eqnarray*}
\rho_{1,\max} (\textbf{X},\textbf{Y}) &=& \frac{ \int_{[0,1]^4} \left( C(u,v) - \Pi (u) W(v) \right) \, du \,dv}{\sqrt{ \var \pi_2(\textbf{U}) \var \pi_2 (\textbf{V}) }} \nonumber\\
& \le & \frac{\frac{767}{13824} - \frac{1}{4}\frac{1}{6}}{\sqrt{ \frac{1}{30} - \frac{1}{36}} \sqrt{ \frac{1}{9} - \frac{1}{16}}} \nonumber\\
& \approx & 0.8408
\end{eqnarray*}
\end{Exa}
Note that similar examples can be given for $\rho_2.$
\end{enumerate}

\subsection{Rank correlation based measures of association}
Instead of applying Pearson correlation to $\pi _{p}\left( \mathbf{U}\right)
$ and $\pi _{q}\left( \mathbf{V}\right) $ as in $\rho_1$ or to $A(\mathbf{U})$ and $B(\mathbf{V})$ as in $\rho_2$ one may apply rank correlation, leading to the measures $\rho_3$ and $\rho_4$.
Let
\begin{align*}
K_{\pi _{p}\left( \mathbf{U}\right) }(t)&:=P\left( \pi _{p}\left( \mathbf{U}%
\right) \leqslant t\right)\,\,\,t\in [0,1], \\
K_{\pi _{q}\left( \mathbf{V}\right) }(t)&:=P\left( \pi _{q}\left( \mathbf{V}
\right) \leqslant t\right)\,\,\,t\in [0,1] \end{align*}
and \begin{align*}
K_{A\left( \mathbf{U}\right)}(t)&:=P\left( A\left( \mathbf{U}\right) \leqslant t\right)\,\,\,t\in \left[ 0,1\right], \\
K_{B\left( \mathbf{V}\right)}(t)&:=P\left( B\left( \mathbf{V}\right) \leqslant t\right)\,\,\,t\in \left[ 0,1\right]
\end{align*}
and let $\ Z_{\pi _{p}\left( \mathbf{U}\right)} :=K_{\pi_{p}\left(\mathbf{U}\right)} \left(\pi_{p}\left( \mathbf{U}\right) \right)$ \ and $Z_{\pi_{q}\left(\mathbf{V}\right)} :=K_{\pi_{q}\left(\mathbf{V}\right)}\left(\pi_{q}\left(\mathbf{V}\right)\right)$,
$Z_{A(\mathbf{U})}:=K_{A(\mathbf{U})}\left( A\left( \mathbf{U}\right) \right) $ and $Z_{B(\mathbf{V})}:=K_{B(\mathbf{V})}\left( B\left( \mathbf{V}\right) \right)$.
From $Z_{\pi _{p}\left( \mathbf{U}\right) },Z_{\pi _{q}\left(
\mathbf{V}\right) }\sim U\left[ 0,1\right] $ it follows that
\begin{align*}
C_{Z_{\pi _{p}\left( \mathbf{U}\right) },Z_{\pi _{q}\left( \mathbf{V}\right)
}}\left( s,t\right) &=P\left( Z_{\pi _{p}\left( \mathbf{U}\right) }\leqslant
s,Z_{\pi _{q}\left( \mathbf{V}\right) }\leqslant t\right) \\
 & =\int_{\left[0,1\right]^p} \int_{\left[0,1\right]^q} \mathds{1}_{\left\{
 K_{\pi_p(\mathbf{U})} (\pi_p(\mathbf{u}))
 \leqslant s\right\} }\mathds{1}_{\left\{
K_{\pi_q(\mathbf{V})} (\pi_q(\mathbf{v}))
\leqslant t\right\} }dC\left( \mathbf{u},\mathbf{v}\right)
\end{align*}
is the joint distribution function and copula of \ $Z_{\pi _{p}\left(
\mathbf{U}\right) }$ \ and \ $Z_{\pi _{q}\left( \mathbf{V}\right) }$.
Implicitly, since
$Z_{A(\mathbf{U})},Z_{B(\mathbf{V})}\sim U\left[ 0,1\right],$ the copula
\begin{align*}
C_{Z_{A(\mathbf{U})},Z_{B(\mathbf{V})}}(s,t)&=P\left( Z_{A(\mathbf{U})}\leqslant s,Z_{B(\mathbf{V})}\leqslant t\right) \\
 &= \int_{\left[0,1\right]^p} \int_{\left[0,1\right]^q} \mathds{1}%
_{\left\{ K_{A(\mathbf{U})} (A(\mathbf{u})) \leqslant s\right\} }%
\mathds{1}_{\left\{ K_{B(\mathbf{V})} (B(\mathbf{v})) \leqslant
t\right\} }dC\left( \mathbf{u},\mathbf{v}\right)
\end{align*}%
is the distribution function and copula of $\left( Z_{A(\mathbf{U})},Z_{B(\mathbf{V})}\right) .$
Based on these copulas, we define the measures
\begin{equation*}
\rho _{3} (C) =\rho _{Z_{\pi _{p}\left( \mathbf{U}\right) },Z_{\pi _{q}\left(
\mathbf{V}\right) }}=12 \int_{0}^{1} \int_{0}^{1} C_{Z_{\pi _{p}\left( \mathbf{U}\right) },Z_{\pi _{q}\left(
\mathbf{V}\right) }}\left( s,t\right) \,dt \,ds -3,
\end{equation*}%
i.e., Spearman's rho of $\ Z_{\pi _{p}\left( \mathbf{U}\right) }$  and
 $Z_{\pi _{q}\left( \mathbf{V}\right)}$ and
\begin{equation*}
\rho _{4} (C) =12 \int_{0}^{1} \int_{0}^{1} %
C_{Z_{A(\mathbf{U})},Z_{B(\mathbf{V})}}(s,t)\,dtds-3,
\end{equation*}%
i.e., Spearman's rho of $Z_{A(\mathbf{U})}$ and $Z_{B(\mathbf{V})}$.
Note that - contrary to the case in section \ref{sec.pearsmeas} - normalisation is not necessary as it is impicitly ensured.
\\
Properties of $\rho _{3}$ and $\rho _{4}:$
\begin{enumerate}
\item We have $-1\leqslant \rho _{3}\leqslant +1$ and $-1\leqslant \rho _{4}\leqslant +1$ and the measures are invariant with respect to permutations within $\mathbf{X}$ and $\mathbf{Y}$.
\item If $\mathbf{X}$ and $\mathbf{Y}$ are independent then $C_{Z_{\pi
_{p}\left( \mathbf{U}\right) },Z_{\pi _{q}\left( \mathbf{V}\right) }}\left(
s,t\right) =C_{Z_{A(\mathbf{U})},Z_{B(\mathbf{V})}}(s,t)=st$ and thus $\rho _{3}=\rho _{4}=0$. Again, the converse is not true.
\item $\rho _{3}=1$ \ is equivalent to
\begin{equation*}
C_{Z_{\pi _{p}\left( \mathbf{U}\right) },Z_{\pi _{q}\left( \mathbf{V}\right)
}}\left( s,t\right) =\min \left\{ s,t\right\}
\end{equation*}
and%
\begin{equation*}
P\left( K_{\pi _{p}\left( \mathbf{U}\right) }\left( \pi _{p}\left( \mathbf{U}%
\right) \right) =K_{\pi _{q}\left( \mathbf{V}\right) }\left( \pi _{q}\left(
\mathbf{V}\right) \right) \right) =1.
\end{equation*}

$\rho _{3}=-1$ \ is equivalent to
\begin{equation*}
C_{Z_{\pi _{p}\left( \mathbf{U}\right) },Z_{\pi _{q}\left( \mathbf{V}\right)
}}\left( s,t\right) =\max \left\{ s+t-1,0\right\}
\end{equation*}

and%
\begin{equation*}
P\left( K_{\pi _{p}\left( \mathbf{U}\right) }\left( \pi _{p}\left( \mathbf{U}%
\right) \right) =1-K_{\pi _{q}\left( \mathbf{V}\right) }\left( \pi
_{q}\left( \mathbf{V}\right) \right) \right) =1.
\end{equation*}
\item $\rho_4=1$ \ is equivalent to \ $C_{Z_{A(\mathbf{U})},Z_{B(\mathbf{V})}}(s,t)=%
\min \left\{ s,t\right\} $
and%
\begin{equation*}
P\left( K_{_{A\left( \mathbf{U}\right)}}\left( A\left( \mathbf{U}%
\right) \right) =K_{B\left( \mathbf{V}\right) }\left( B\left(
\mathbf{V}\right) \right) \right) =1.
\end{equation*}
$\rho _{4}=-1$ \ is equivalent to
\begin{equation*}
C_{Z_{A\left( \mathbf{U}\right) },Z_{B\left( \mathbf{V}\right)
}}\left( s,t\right) =\max \left\{ s+t-1,0\right\}
\end{equation*}
and%
\begin{equation*}
P\left( K_{A\left( \mathbf{U}\right) }\left(A\left( \mathbf{U}%
\right) \right) =1-K_{B\left( \mathbf{V}\right) }\left(B\left( \mathbf{V}\right) \right) \right) =1.
\end{equation*}
\item Contrary to $\rho _{1}$ and $\rho _{2}$, the rank correlation based measures $\rho _{3}$ and $\rho _{4}$
can achieve every value between $-1$ and $+1$ for any given and fixed marginal
copulas $A$ and $B$.
To show this for $\rho _{3}$ and $+1$
 let $(S, T)$ be a comonotone pair of random variables such that S has the same distribution as $\pi_p(\mathbf{U})$ and $T$ has the same distribution as $\pi_q(\mathbf{V})$. Second, conditionally on $(S, T) = (s, t),$ let $\mathbf{U}'$ and $\mathbf{V}'$ be independent random vectors whose conditional distributions are given by the one of $\mathbf{U}$ given $\pi_p(\mathbf{U}) = s$ and the one of $\mathbf{V}$ given $\pi_q(\mathbf{V}) = t,$ respectively. Now, the copulas of $\mathbf{U}'$ and $\mathbf{V}'$ are $A$ and $B$ by construction and $\rho_3=1$ due to the comonotonicity of $(S, T).$
Analogous arguments hold for $\rho _{3}=-1$ and $\rho_4=\pm 1.$
\end{enumerate}

\section{Statistical estimation of the measures}
\label{sec.empest}
In this section we propose nonparametric estimators of the discussed measures.
It is assumed that the marginal distribution functions $F_{X_i}$ and $F_{Y_j}$ are unknown for $i=1,\ldots,p$ and $j=1,\ldots,q.$

Let $\left( \mathbf{X}_{1},\mathbf{Y}_{1}\right) ,\ldots,\left( \mathbf{X}_{n},%
\mathbf{Y}_{n}\right) $ be i.i.d. samples from $\left( \mathbf{X},\mathbf{Y}\right).$
Let $\widehat{F}_{X_{i},n}$ and $\widehat{F}_{Y_{j},n}$ \ denote the
empirical distribution function of $X_{i}$ and $Y_{j}$ for \ $i=1,\ldots,p$ \
and\ \ $j=1,\ldots,q.$ Then, for $k=1,\ldots,n,$%
$$
\widehat{\mathbf{U}}_{k,n}=\left( \widehat{F}_{X_{1},n}\left( X_{1k}\right)
,\ldots,\widehat{F}_{X_{p},n}\left( X_{pk}\right) \right) $$
and
$$\widehat{\mathbf{V}}_{k,n}=\left( \widehat{F}_{Y_{1},n}\left( Y_{1k}\right)
,\ldots,\widehat{F}_{Y_{q},n}\left( Y_{qk}\right) \right) $$%
are called the pseudo observations of%
$$\mathbf{U}_{k}=\left( F_{X_{1}}\left( X_{1k}\right) ,\ldots,F_{X_{p}}\left(
X_{pk}\right) \right)$$
and
$$\mathbf{V}_{k}=\left( F_{Y_{1}}\left( Y_{1k}\right) ,\ldots,F_{Y_{q}}\left(
Y_{qk}\right) \right).$$
The empirical distribution function of $\left( \widehat{\mathbf{U}}_{k,n},%
\widehat{\mathbf{V}}_{k,n}\right) $ for $\ k=1,\ldots,n$, i.e.,
\begin{equation*}
\widehat{C}_{n}\left( \mathbf{u},\mathbf{v}\right) = \frac{1}{n}
\sum_{k=1}^{n} \mathbf{1}_{\left\{ \widehat{\mathbf{U}}_{k,n}\leq
\mathbf{u}\right\} }\mathbf{1}_{\left\{ \widehat{\mathbf{V}}_{k,n}\leq
\mathbf{v}\right\} }
\end{equation*}%
is the empirical copula of $\left( \mathbf{X},\mathbf{Y}\right)$ (see, e.g., \citet{Deheuvels1979}).
The marginal empirical copulas of $\mathbf{X}$ and $\mathbf{Y}$ are estimated by
\begin{equation*}
\widehat{A}_{n}\left( \mathbf{u}\right) =\widehat{C}_{n}\left( \mathbf{u},%
\mathbf{1}_{q}\right) \text{ \ \ and \ \ }\widehat{B}_{n}\left( \mathbf{v}%
\right) =\widehat{C}_{n}\left( \mathbf{1}_{p},\mathbf{v}\right) .
\end{equation*}%
In the following, estimation is based on the pseudo observations
\begin{equation*}
\left( \widehat{\mathbf{U}}_{1,n},\ldots,\widehat{
\mathbf{U}}_{n,n}\right) \text{ \ and \ }\left(
\widehat{\mathbf{V}}_{1,n},\ldots,\widehat{\mathbf{V}}_{n,n}\right).
\end{equation*}

\textbf{Estimation of }$\overline{\rho}_{\mathbf{X},\mathbf{Y}}$.
The estimator for $\overline{\rho }_{\mathbf{X},\mathbf{Y}}$ is
given by
\begin{eqnarray*}
\widehat{\overline{\rho}}_n &=& \overline{\rho} \left( \widehat{C}_n \right)
 =  \frac{1}{pq} \sum_{i=1}^{p} \sum_{j=1}^{q} \left( 12
\int_{\left[0,1\right]^2} \widehat{C}_{ij,n}\left( u_{i},v_{j}\right) \, du_{i} \, dv_{j}-3\right) \\
& =&   \frac{1}{pq} \sum_{i=1}^{p} \sum_{j=1}^{q} \left( 12\left( \frac{1}{n}
\sum_{k=1}^{n} \left( 1-\widehat{U}_{ki,n}\right) \left( 1-\widehat{V}%
_{kj,n}\right) \right) -3\right) .
\end{eqnarray*}
%

\textbf{Estimation of }$\rho _{1}$.
The estimator $\widehat{\rho }_{1,n}$ for $\rho _{1}$ is the Pearson coefficient of correlation of
\begin{equation*}
\pi _{p}\left( \widehat{\mathbf{U}}_{1,n}\right) ,\ldots,\pi _{p}\left(
\widehat{\mathbf{U}}_{n,n}\right) \text{ \ and \ }\pi _{q}\left( \widehat{
\mathbf{V}}_{1,n}\right) ,\ldots,\pi _{p}\left( \widehat{\mathbf{V}}
_{n,n}\right).
\end{equation*}

\textbf{Estimation of }$\rho _{2}$.
We first estimate $A$ and $B$ by
\begin{equation*}
\widehat{A}_{n}\left( \mathbf{u}\right) =\widehat{C}_{n}\left( \mathbf{u},%
\mathbf{1}_{q}\right) \text{ \ and \ }\widehat{B}_{n}\left( \mathbf{v}%
\right) =\widehat{C}_{n}\left( \mathbf{1}_{p},\mathbf{v}\right)
\end{equation*}%
and obtain pseudo observations on $A\left( \mathbf{U}\right) $ and $B\left(
\mathbf{V}\right) $ by%
\begin{equation*}
\widehat{A}_{n}\left( \widehat{\mathbf{U}}_{1,n}\right) ,\ldots,\widehat{A}_{n}\left(
\widehat{\mathbf{U}}_{n,n}\right) \text{ \ and \ }\widehat{B}_{n}\left( \widehat{\mathbf{V}}%
_{1,n}\right) ,\ldots,\widehat{B}_{n}\left( \widehat{\mathbf{V}}_{n,n}\right) .
\end{equation*}%
$\widehat{\rho }_{2,n}$ is the Pearson coefficient of correlation of these pseudo
observations.

\textbf{Estimation of }$\rho _{3}$.
In order to estimate $\rho _{3},$ we first have to estimate $K_{\pi
_{p}\left( \mathbf{U}\right) }$ by
\begin{equation*}
\widehat{K}_{\pi _{p}\left( \mathbf{U}\right) ,n}\left( t\right) =\frac{1}{n}%
\sum_{k=1}^{n} \mathds{1}_{\left\{ \pi _{p}\left(
\widehat{\mathbf{U}}_{k,n}\right) \leqslant t\right\} }
\end{equation*}%
for \ $t\in \left[ 0,1\right] $ \ and similarly $K_{\pi _{q}\left( \mathbf{V}%
\right) }$. Let%
\begin{equation*}\widehat{Z}_{\pi _{p}\left( \mathbf{U}\right), k,n}=\widehat{K}_{\pi
_{p}\left( \mathbf{U}\right) ,n}\left( \pi _{p}\left( \widehat{\mathbf{U}}%
_{k,n}\right) \right) =\frac{1}{n} \sum_{l=1}^{n} \mathds{1}_{\left\{ \pi _{p}\left( \widehat{\mathbf{U}}_{l,n}\right)
\leqslant \pi _{p}\left( \widehat{\mathbf{U}}_{k,n}\right) \right\} }\end{equation*}%
and
\begin{equation*}\widehat{Z}_{\pi _{q}\left( \mathbf{V}\right), k,n}=\widehat{K}_{\pi
_{q}\left( \mathbf{V}\right) ,n}\left( \pi _{q}\left( \widehat{\mathbf{V}}%
_{k,n}\right) \right) =\frac{1}{n} \sum_{l=1}^{n} \mathds{1}_{\left\{ \pi _{q}\left( \widehat{\mathbf{V}}_{l,n}\right)
\leqslant \pi _{q}\left( \widehat{\mathbf{V}}_{k,n}\right) \right\} }\end{equation*}%
for $k=1,\ldots,n$. Then with
\begin{equation*}
\widehat{C}_{Z_{\pi _{p}\left( \mathbf{U}\right) },Z_{\pi _{q}\left( \mathbf{V}%
\right) ,n}}\left( s,t\right) :=\frac{1}{n}\sum_{k=1}^{n}%
\mathds{1}_{\left\{ \widehat{Z}_{\pi _{p}\left( \mathbf{U}\right), k,n}
\leqslant s\right\} }\mathds{1}_{\left\{ \widehat{Z}_{\pi _{q}\left( \mathbf{V}\right), k,n} \leqslant t\right\} }
\end{equation*}%
an estimator for $\rho_3$ is
\begin{eqnarray*}
\widehat{\rho }_{3,n} & = & 12\int_{0}^{1}\int_{0}^{1} \widehat{C}_{\pi_{p}\left( \mathbf{U}%
\right) ,\pi_{q}\left( \mathbf{V}\right) ,n}(s,t) \, dtds-3 \\
& = & 12\frac{1}{n}\sum_{k=1}^{n}\left( 1-\widehat{Z}%
_{\pi_{p}\left( \mathbf{U}\right) ,k,n}\right) \left( 1-\widehat{Z}_{\pi_{q}\left( \mathbf{V}\right) ,k,n}\right) -3.
\end{eqnarray*}

\textbf{Estimation of }$\rho _{4}$.
The estimator $\widehat{\rho}_{4,n}$ for $\rho _{4}$ is derived in a similar way as $\widehat{\rho}_{3,n}.$
Let
\begin{equation*}
\widehat{K}_{A(\mathbf{U}),n}\left( s\right) = \frac{1}{n}\sum_{k=1}^{n} \mathds{1}_{\left\{ \widehat{A}_{n}\left( \widehat{U}_{k,n}\right) \leqslant s\right\}}
\end{equation*}
and
\begin{equation*}
\widehat{K}_{B(\mathbf{V}),n}\left( t\right) = \frac{1}{n}\sum_{k=1}^{n}\mathds{1}_{\left\{ \widehat{B}_{n}\left( \widehat{V}_{k,n}\right)
\leqslant t\right\} }.
\end{equation*}
Then
\begin{equation*}
\widehat{K}_{A(\mathbf{U}),n}\left( \widehat{A}_{n}\left( \widehat{%
\mathbf{U}}_{k,n}\right) \right) =\widehat{Z}_{A(\mathbf{U}),k,n} = \frac{1}{n} \sum_{l=1}^{n} \mathds{1}_{\left\{ \widehat{A}_{n}\left( \widehat{\mathbf{U}}_{l,n}\right) \leqslant
\widehat{A}_{n}\left( \widehat{\mathbf{U}}_{k,n}\right) \right\} }
\end{equation*}
and
\begin{equation*}
\widehat{K}_{B(\mathbf{V}),n}\left( \widehat{B}_{n}\left( \widehat{\mathbf{V}}
_{k,n}\right) \right) = \widehat{Z}_{B(\mathbf{V}),k,n} =\frac{1}{n} \sum_{l=1}^{n} \mathds{1}%
_{\left\{ \widehat{B}_{n}\left( \widehat{\mathbf{V}}_{l,n}\right) \leqslant
\widehat{B}_{n}\left( \widehat{\mathbf{V}}_{k,n}\right) \right\} }
\end{equation*}
for $k=1,\ldots,n$.
For
\begin{eqnarray*}
\widehat{C}_{Z_{A(\mathbf{U})},Z_{B(\mathbf{V})},n}\left( s,t\right) :=\frac{1}{n}\sum_{k=1}^{n} \mathds{1}_{\left\{ \widehat{Z}_{A(\mathbf{U}),k,n}\leqslant s\right\} }\mathds{1}_{\left\{ \widehat{Z}_{B(\mathbf{V}),k,n}\leqslant t\right\} }
\end{eqnarray*}
it follows
\begin{eqnarray*}
\int_{0}^{1}\int_{0}^{1} \widehat{C}_{Z_{A(\mathbf{U})},Z_{B,n}}\left( s,t\right) dt ds = \frac{1}{%
n}\sum_{k=1}^{n}\left( 1-\widehat{Z}_{A(\mathbf{U}),k,n}\right)
\left( 1-\widehat{Z}_{B(\mathbf{V}),k,n}\right),
\end{eqnarray*}
and thus
\begin{equation*}
\widehat{\rho}_{4} = 12\left( \frac{1}{n} \sum_{k=1}^{n}\left( 1-\widehat{Z}_{A(\mathbf{U}),k,n}\right) \left(
1-\widehat{Z}_{B(\mathbf{V}),k,n}\right) \right) -3.
\end{equation*}%
We have derived asymptotic normality of $\widehat{\overline{\rho}}_n$ and $\widehat{\rho}_1$ using the asymptotic theory of the copula process introduced by \citet{rueschendorf76} (see, e.g., \citet{Fermanian2004} and \citet{segers10} for recent references) and the functional delta method (see, e.g., \citet{vaartwellner96}). Details can be obtained from the authors. The asymptotic normality of $\widehat{\rho}_{i,n}$ for $i=2,3,4$ has not yet been derived.

In the following, the finite sample properties of $\widehat{\overline{\rho}}_n$ and $\widehat{\rho}_{1,n}, \ldots, \widehat{\rho}_{4,n}$,
i.e. their bias and standard deviation are investigated by a Monte Carlo simulation. In order to estimate the standard deviation of the estimators, we use the bootstrap and the jackknife.

For a given sample $(\textbf{X}_1,\textbf{Y}_1),\ldots, (\textbf{X}_n,\textbf{Y}_n)$ of i.i.d. observations, the bootstrap draws $n$ observations of the sample with replacement. Ties are solved by mid-ranks.
 For $B$ bootstrap samples, the standard deviation is estimated by
\begin{equation*}
\hat{\sigma}_{\widehat{\rho}_n}^B = \sqrt{ \frac{1}{B-1} \sum_{b=1}^B \left( \hat{\rho}_n^{(b)} - \overline{\hat{\rho}}_n \right)^2},
\end{equation*}
where $\hat{\rho}_n^{(b)}$ is the estimator of the association of the $b$-th bootstrap sample and $\overline{\hat{\rho}}_n$
their mean.

For the jackknife estimator of the standard deviation of $\hat{\rho}_n$ let $\hat{\rho}_{n-1}^{(j)}$ denote the estimator, where the $j$-th observation of $(\textbf{X}_1,\textbf{Y}_1), \ldots, (\textbf{X}_n,\textbf{Y}_n)$ is deleted.
The jackknife
estimate of the standard deviation is then given by
\begin{equation*}
\hat{\sigma}_{\widehat{\rho}_n}^J = \sqrt{ \frac{n-1}{n} \sum_{j=1}^{n} \left( \hat{\rho}_{n-1}^{(j)} - \overline{\hat{\rho}}_{n-1} \right)^2}.
\end{equation*}

For the simulation study, we consider observations from the Gaussian copula (see \cite{Joe1997}) and the Clayton copula (see \cite{Clayton1978})
with different dimensions and different sample sizes. The Gaussian copula is defined by
\begin{equation*}
	C_{\Theta} (u_1,\ldots,u_d) := \Phi_{\Theta} \left( \Phi^{-1} (u_1),\ldots, \Phi^{-1} (u_d) \right),
\end{equation*}
where $\Phi_{\Theta}$ is the distribution function of the multivariate normal distribution with zero mean, unit variances and positive definite correlation matrix $\Theta = (\theta_{ij})_{i,j=1,\dots,d}$. Further, $\Phi^{-1}$ denotes the quantile function of the univariate standard normal distribution.

The $d$-dimensional Clayton copula (see \cite{Clayton1978}) is given by
\begin{equation*}
	C_{\theta} (u_1,\ldots, u_n) := \left( \sum_{i=1}^d u_i^{-\theta} - d + 1 \right)^{-1/\theta},
\end{equation*}
for $\theta>0$.

To reduce the number of parameters in our model, we only consider the case of equi-correlation for the Gaussian copula, although simulations
with more complex correlation matrices show similar results. The results are based
on 10,000 Monte Carlo simulations and $500$ bootstrap iterations, respectively.

Tables \ref{table:bootstrap-gaussian_rho_bar} to \ref{table:bootstrap-gaussian_rho_4} show simulation results with the Gaussian copula for $\overline{\rho}$, $\rho_2$ and $\rho_4$ for two random vectors of dimensions $p=q=3$ and $p=q=4$ and sample sizes $50,$ $100$ and $500$ as well as different correlation parameters $\theta$. Results for the remaining measures and for the Clayton copula are omitted, but can be obtained
from the authors. They are, however, very similar to the results presented.

The first two columns in the tables contain the value of the dependence parameters and the sample sizes. The third column of the tables shows
an approximation to the true value of the measures of association, which has been derived from samples
of size 1,000,000. Comparing the true values to the mean of the estimated associations $m(\hat{\rho}_n)$ in column 4, we observe a small finite sample
bias, which decreases with increasing sample size.
The standard deviation of the estimator $s(\hat{\rho}_n)$ and the means of the bootstrap estimation $m(\hat{\sigma}^B)$ and the
jackknife estimation $m(\hat{\sigma}^J)$ are shown in colums $5,6$ and $7$. It can be seen that both procedures for the estimation of the standard
deviation perform well for the Gaussian copula. Furthermore, the standard deviation of the estimator decreases
with increasing sample size in a reasonable way. Finally, columns $8$ and $9$ show that the standard error of the bootstrap standard deviation estimates
is slightly smaller than the obtained jackknife estimates.

\begin{table}[!ht]
\centering
\begin{tabular}{*{9}{c}} \toprule
$\theta$	&	n		&	$\overline{\rho}$	&	$m(\widehat{\overline{\rho}}_{n})$ & $s(\widehat{\overline{\rho}}_{n})$ &	$m(\hat{\sigma}^B)$ & $m(\hat{\sigma}^J)$ & $s(\hat{\sigma}^B)$ & $s(\hat{\sigma}^J)$ \\\midrule
\multicolumn{9}{l}{Two 3-dimensional vectors}
\\\midrule
-0.1 & 50  & -0.096 & -0.094 & 0.040 & 0.039 & 0.040 & 0.007 & 0.007 \\
     & 100 & -0.096 & -0.095 & 0.027 & 0.028 & 0.028 & 0.003 & 0.003 \\
     & 500 & -0.096 & -0.096 & 0.012 & 0.012 & 0.012 & 0.001 & 0.001 \\
0.2  & 50  & 0.191  & 0.188  & 0.063 & 0.063 & 0.064 & 0.008 & 0.008 \\
     & 100 & 0.191  & 0.189  & 0.045 & 0.044 & 0.045 & 0.004 & 0.004 \\
     & 500 & 0.191  & 0.191  & 0.020 & 0.020 & 0.020 & 0.001 & 0.001 \\
0.5  & 50  & 0.483  & 0.474  & 0.069 & 0.070 & 0.071 & 0.007 & 0.008 \\
     & 100 & 0.483  & 0.479  & 0.049 & 0.049 & 0.049 & 0.004 & 0.004 \\
     & 500 & 0.483  & 0.482  & 0.022 & 0.022 & 0.022 & 0.001 & 0.001 \\
\midrule	
\multicolumn{9}{l}{Two 4-dimensional vectors}
\\\midrule
-0.1 & 50  & -0.096 & -0.094 & 0.028 & 0.027 & 0.028 & 0.005 & 0.005 \\
     & 100 & -0.096 & -0.095 & 0.019 & 0.019 & 0.020 & 0.003 & 0.003 \\
     & 500 & -0.096 & -0.095 & 0.009 & 0.009 & 0.009 & 0.001 & 0.001 \\
0.2  & 50  & 0.191  & 0.188  & 0.054 & 0.054 & 0.055 & 0.007 & 0.007 \\
     & 100 & 0.191  & 0.189  & 0.038 & 0.038 & 0.039 & 0.004 & 0.004 \\
     & 500 & 0.191  & 0.191  & 0.017 & 0.017 & 0.017 & 0.001 & 0.001 \\
0.5  & 50  & 0.483  & 0.474  & 0.064 & 0.064 & 0.065 & 0.006 & 0.007 \\
     & 100 & 0.483  & 0.478  & 0.045 & 0.045 & 0.046 & 0.003 & 0.003 \\
     & 500 & 0.483  & 0.481  & 0.020 & 0.020 & 0.020 & 0.001 & 0.001 \\
\bottomrule									
\end{tabular}
\caption{Simulation results for a equicorrelated Gaussian copula with correlation $\theta$ and sample size $n$. $\overline{\rho}$ denotes the theoretical value. The empirical means $m(\cdot)$ and the empirical standard deviations $s(\cdot)$ are based on 10,000 Monte Carlo simulations and $500$ bootstrap samples. The bootstrap estimates are labeled by the superscript B, the jackknife estimates by $J$.}
\label{table:bootstrap-gaussian_rho_bar}
\end{table}


\begin{table}[!ht]
\centering
\begin{tabular}{*{9}{c}} \toprule
$\theta$	&	n		&	$\rho_2$	&	$m(\hat{\rho}_{2,n})$ & $s(\hat{\rho}_{2,n})$ &	$m(\hat{\sigma}^B)$ & $m(\hat{\sigma}^J)$ & $s(\hat{\sigma}^B)$ & $s(\hat{\sigma}^J)$ \\\midrule
\multicolumn{9}{l}{Two 3-dimensional vectors}
\\\midrule
-0.1 & 50  & -0.225 & -0.218 & 0.104 & 0.113 & 0.110 & 0.029 & 0.038 \\
     & 100 & -0.225 & -0.220 & 0.071 & 0.072 & 0.073 & 0.018 & 0.021 \\
     & 500 & -0.225 & -0.225 & 0.031 & 0.031 & 0.031 & 0.005 & 0.005 \\
0.2  & 50  & 0.349  & 0.335  & 0.146 & 0.143 & 0.155 & 0.021 & 0.030 \\
     & 100 & 0.349  & 0.343  & 0.103 & 0.102 & 0.106 & 0.012 & 0.015 \\
     & 500 & 0.349  & 0.347  & 0.046 & 0.046 & 0.046 & 0.003 & 0.003 \\
0.5  & 50  & 0.682  & 0.664  & 0.095 & 0.098 & 0.100 & 0.021 & 0.025 \\
     & 100 & 0.682  & 0.673  & 0.066 & 0.066 & 0.067 & 0.012 & 0.012 \\
     & 500 & 0.682  & 0.681  & 0.029 & 0.029 & 0.029 & 0.003 & 0.002 \\
\midrule	
\multicolumn{9}{l}{Two 4-dimensional vectors}
\\\midrule
-0.1 & 50  & -0.235 & -0.226 & 0.073 & 0.123 & 0.087 & 0.025 & 0.033 \\
     & 100 & -0.235 & -0.231 & 0.048 & 0.058 & 0.054 & 0.013 & 0.015 \\
     & 500 & -0.235 & -0.234 & 0.021 & 0.021 & 0.021 & 0.003 & 0.003 \\
0.2  & 50  & 0.388  & 0.370  & 0.157 & 0.151 & 0.168 & 0.025 & 0.045 \\
     & 100 & 0.388  & 0.378  & 0.110 & 0.108 & 0.115 & 0.016 & 0.023 \\
     & 500 & 0.388  & 0.387  & 0.050 & 0.049 & 0.050 & 0.004 & 0.004 \\
0.5  & 50  & 0.723  & 0.666  & 0.094 & 0.097 & 0.099 & 0.021 & 0.026 \\
     & 100 & 0.723  & 0.674  & 0.065 & 0.066 & 0.067 & 0.011 & 0.012 \\
     & 500 & 0.723  & 0.721  & 0.028 & 0.028 & 0.028 & 0.003 & 0.003 \\
\bottomrule	
\end{tabular}
\caption{Simulation results for a equicorrelated Gaussian copula with correlation $\theta$ and sample size $n$. $\rho_2$ denotes the approximation of the theoretical value, estimated from a sample with sample size 100,000,000. The empirical means $m(\cdot)$ and the empirical standard deviations $s(\cdot)$ are based on 10,000 Monte Carlo simulations and $500$ bootstrap samples. The bootstrap estimates are labeled by the superscript B, the jackknife estimates by $J$.}
\label{table:bootstrap-gaussian_rho_2}
\end{table}


\begin{table}[!ht]
\centering
\begin{tabular}{*{9}{c}} \toprule
$\theta$	&	n		&	$\rho_4$	&	$m(\hat{\rho}_{4,n})$ & $s(\hat{\rho}_{4,n})$ &	$m(\hat{\sigma}^B)$ & $m(\hat{\sigma}^J)$ & $s(\hat{\sigma}^B)$ & $s(\hat{\sigma}^J)$ \\\midrule
\multicolumn{9}{l}{Two 3-dimensional vectors}
\\\midrule
-0.1 & 50  & -0.310 & -0.250 & 0.126 & 0.133 & 0.147 & 0.010 & 0.017 \\
     & 100 & -0.310 & -0.277 & 0.091 & 0.089 & 0.100 & 0.006 & 0.008 \\
     & 500 & -0.310 & -0.302 & 0.041 & 0.039 & 0.042 & 0.002 & 0.001 \\
0.2  & 50  & 0.375  & 0.330  & 0.123 & 0.118 & 0.135 & 0.012 & 0.016 \\
     & 100 & 0.375  & 0.353  & 0.086 & 0.083 & 0.092 & 0.007 & 0.008 \\
     & 500 & 0.375  & 0.369  & 0.039 & 0.038 & 0.040 & 0.002 & 0.002 \\
0.5  & 50  & 0.694  & 0.635  & 0.081 & 0.084 & 0.090 & 0.015 & 0.019 \\
     & 100 & 0.694  & 0.665  & 0.057 & 0.057 & 0.060 & 0.008 & 0.010 \\
     & 500 & 0.694  & 0.688  & 0.025 & 0.025 & 0.025 & 0.002 & 0.002 \\
\midrule	
\multicolumn{9}{l}{Two 4-dimensional vectors}
\\\midrule
-0.1 & 50	 & -0.462 & -0.265 & 0.110 &	0.150 &	0.144	& 0.011	& 0.021 \\
     & 100 & -0.462 & -0.337 & 0.085 &	0.101	& 0.102	& 0.007	& 0.010 \\
     & 500 & -0.462 & -0.428 & 0.038 &	0.036	& 0.040	& 0.002	& 0.002 \\
0.2	 & 50	 & 0.433 & 0.356  & 0.118 &	0.116	& 0.137	& 0.012	& 0.017 \\
   	 & 100 & 0.433 & 0.394  & 0.085 &	0.080 &	0.092	& 0.007	& 0.009 \\
   	 & 500 & 0.433 & 0.426  & 0.037 &	0.036 &	0.038	& 0.002	& 0.002 \\
0.5	 & 50	 & 0.740 & 0.667  & 0.077 &	0.080 &	0.086	& 0.015 & 0.020 \\
   	 & 100 & 0.740 & 0.704  & 0.052	& 0.052 &	0.055	& 0.008	& 0.010 \\
   	 & 500 & 0.740 & 0.733  & 0.022 &	0.022	& 0.023	& 0.002	& 0.002 \\
\bottomrule									
\end{tabular}
\caption{Simulation results for a equicorrelated Gaussian copula with correlation $\theta$ and sample size $n$. $\rho_4$ denotes the approximation of the theoretical value, estimated from a sample with sample size 100,000,000. The empirical means $m(\cdot)$ and the empirical standard deviations $s(\cdot)$ are based on 10,000 Monte Carlo simulations and $500$ bootstrap samples. The bootstrap estimates are labeled by the superscript B, the jackknife estimates by $J$.}
\label{table:bootstrap-gaussian_rho_4}
\end{table}


\FloatBarrier

We further investigate how well the finite sample distribution of the introduced empirical measures of association can be approximated by the normal distribution.
To this end, we compute $\bar{\rho}, \rho_1, \rho_2, \rho_3$ and $\rho_4$ for $N=$10,000 Monte Carlo simulations from two $2$-dimensional random vectors from the Gaussian copula and the Clayton copula. For each copula, we use different dependence parameters and various sample sizes. We standardise the 10,000 measures obtained from the Monte Carlo simulation by their sample mean and standard deviation, respectively, and use a kernel estimator to approximate their density.

\begin{figure}[!ht]
	\centering
		\includegraphics[width=1\textwidth]{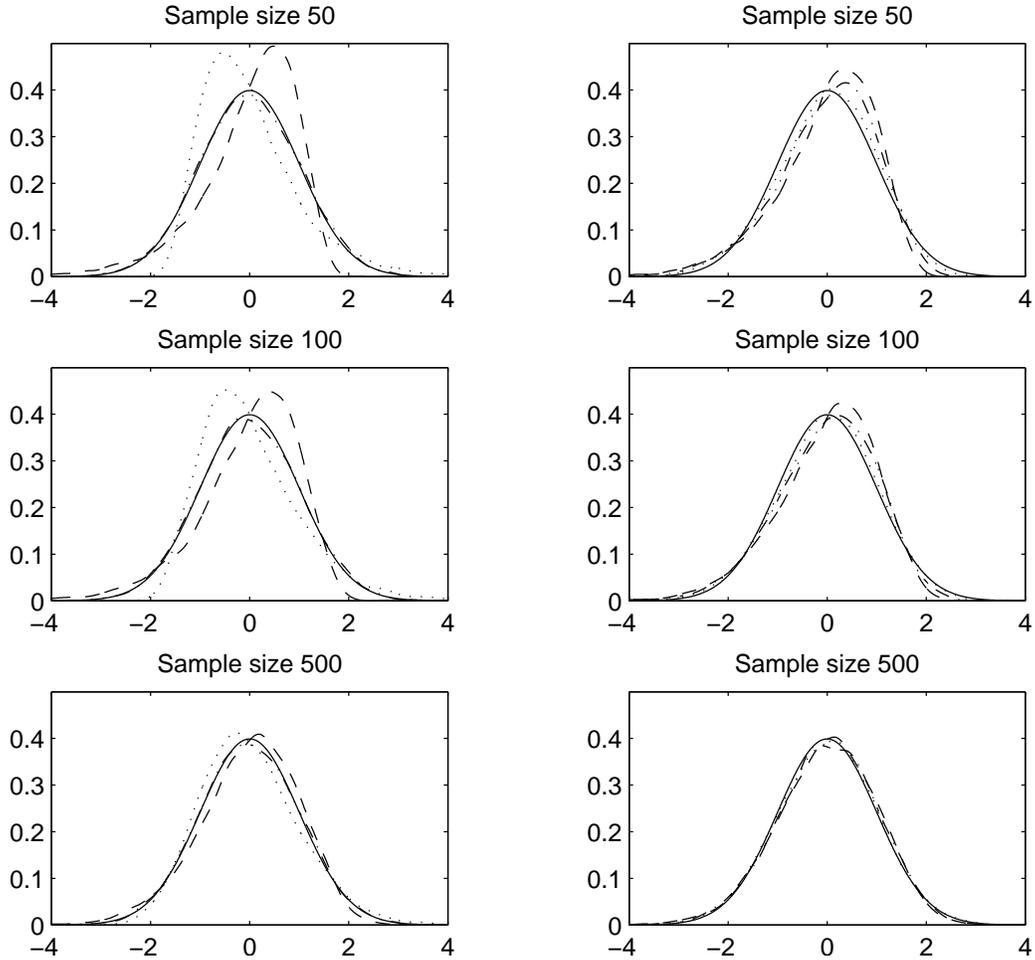}
	\caption{Kernel density estimates of the normalised estimator for $\rho_3$ for the Gaussian copula with $\varrho=-0.75,0$ and $0.75$ (left panel) and the 4 dimensional
					 Clayton Copula with $\alpha=0.5, 2$ and $5$ (right panel). The solid line depicts the standard normal distribution.}
	\label{fig:asymptotics}	
\end{figure}

The left panel of figure \ref{fig:asymptotics} shows the results for $\rho_3$ in case of the Gaussian copula, where the correlation matrix has the form
\[ \Sigma = \begin{pmatrix} 1 & 0.5 & \varrho & \varrho\\ 0.5 & 1 & \varrho & \varrho\\ \varrho & \varrho & 1 & 0.5\\ \varrho & \varrho & 0.5 & 1 \end{pmatrix} \]
with $\varrho=-0.75, 0$ and $0.75$. Whereas the density of the estimator is highly skewed for $\varrho=-0.75$ and $\varrho=0.75$ for a sample size of $50$,
this asymmetry vanishes with increasing sample size and the density of the $\rho_3$ for all of the considered values of $\varrho$ is barely
distinguishable from the normal density for a sample size of 500. It has to be noted that $-0.75$ and $0.75$ are the upper and lower bound of $\varrho$ such that $\Sigma$ is a correlation matrix. For values closer to $0$, the asymmetries are smaller.

The right panel of figure \ref{fig:asymptotics} shows the results for a Clayton Copula with parameters $\alpha=0.5, 2$ and $5$. Again, we used $\rho_3$
to measure the association. The other measures, however, show similar results, which are available upon request by the authors. For a sample
size of $50$, the density of the estimator is slightly skewed for all parameters, nevertheless the highest skewness occurs for $\alpha = 5$.
As for the Gaussian copula, the skewness decreases with increasing sample size and is barely observable for a sample size of 500.

Having performed similar Monte Carlo simulations for other dimensions and dependence parameters, we conclude that for a sample size of $500$
the finite sample distribution of the association measure can very well be approximated by the normal distribution.

\FloatBarrier

\section{Empirical example}
\label{sec.empex}

\begin{figure}[!ht]
	\centering
		\includegraphics[width=1\textwidth]{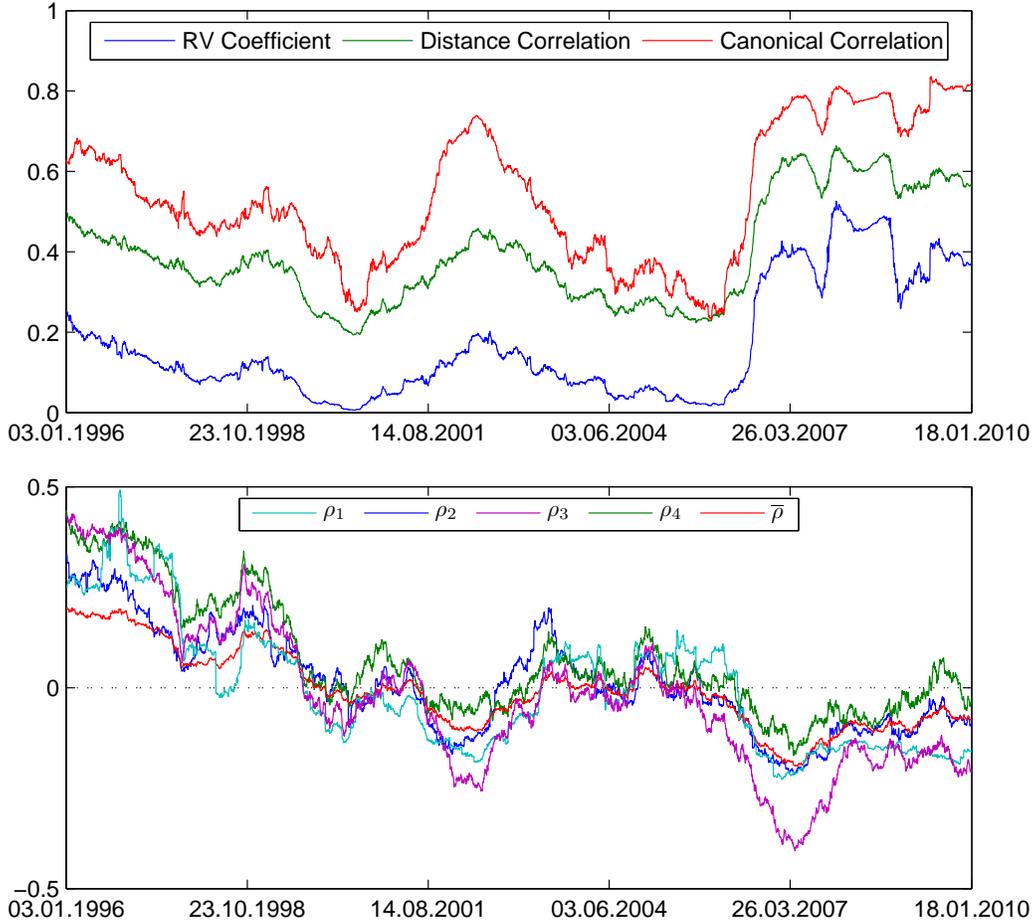}
	\caption{Evolution of the association between the bonds and stock markets, measured by their canonical correlation, their distance correlation and
	their RV coefficient (top panel) and by the introduced measures of association $\overline{\rho}$ and $\rho_1,\ldots,\rho_4$ (bottom panel). The
	analysis is based on a moving window approach	with a window size of 250 days.}
	\label{fig:dependence-plot}	
\end{figure}

In our empirical example we make an attempt to measure strength and direction of association of the bond and the stock market. We make use of the five copula based measures of association presented in section \ref{sec.popver}. Moreover, for the sake of comparison, the traditional canonical correlation, the RV coefficient and distance correlation are applied.
We consider daily returns of the stock market indices of five major countries\footnote{All Ordinaries (Australia), CAC 40 (France),
DAX (Germany), Nikkei 225 (Japan) and S\&P 500 (USA)} as well as government bonds indices from The Bank of America Merrill Lynch \footnote{The BofA Merrill Lynch Australia Government Index (G0T0), The BofA Merrill Lynch French Government Index (G0F0), The BofA Merrill Lynch Japan Government Index (G0Y0), The BofA Merrill Lynch German Government Index (G0D0) and The BofA Merrill Lynch Australia Government Index (G0T0)}
for the respective countries during the period from January 3rd, 1996 to December 30th, 2010. Figure \ref{fig:dependence-plot} shows the evolution of the association of bond and stock market, based on a forward-looking moving window with a window size of 250 days.
In particular, the first value of each measure is based on the 250 daily returns following January, 2rd, 1996. The last value is estimated from the 250 daily returns from January 18th, 2010 until the end of 2010.
The top panel shows that canonical correlation, distance correlation and the RV coefficient exhibit similar patterns of association. Their scaling is quite different, however. Canonical correlation is always highest, RV coefficient always lowest. Distance correlation is somehow between the two, but closer to the canonical correlation in general. The evolution of association over time as indicated by the five measures $\overline{\rho}$ and $\rho_1,\ldots,\rho_4$ is shown in the bottom panel. The differences between their values are in general smaller than these of the aforementioned measures. The evolution of $\overline{\rho}$ seems to be smoothest which is not surprising, while $\rho_3$ is most erratic. It can be seen that there is a tendency of decreasing association from 1996 to 2002, association is close to zero between 2002 and 2006 and becomes negative afterwards. A pattern of association like this can only be recorded by using measures of the type which we introduced in section \ref{sec.popver}. Note that the identified pattern of association is an empirical finding, for which we do not attempt to provide an economic explanation.
The graphs in figure \ref{fig:dependence-plot} are based on the returns themselves. We have, however, made similar graphs for filtered data where autocorrelation and heteroscedasticity have been removed. The results for the filtered data are very close to those of the unfiltered data.

\section{Conclusion}

\label{sec.concl}

Five measures of association between two random vectors $\mathbf{X}$ and $\mathbf{Y}$ have been introduced. They are copula based and do therefore not depend on the marginal distributions of the components $X_i$ and $Y_j.$ They measure strength and direction of association, so they are capable of distinguishing positive and negative association. This is a substantial advantage in applications to real life data, in particular financial data. Estimators for the measures have been proposed and it was demonstrated by simulation that they have favorable small sample properties at least for $n\geq 100.$ There is space for extension and complementation of the measures. First, it can be seen that the measures $\rho_1, \dots, \rho_4$ have in common that they are based on transformations $f(\mathbf{U})$ and $g(\mathbf{V})$, say, where $f$ and $g$ may, or may not, depend on the marginal copulas $A$ and $B.$ Therefore more general classes of measures can be defined, if further measures of bivariate association are applied to $f(\mathbf{U})$ and $g(\mathbf{V}).$

Second, measures for association between $d$ vectors $\mathbf{X}_1, \mathbf{X}_2, \mathbf{X}_3 \dots, \mathbf{X}_d$ of dimensions $p_1,p_2,\dots,p_d,$ respectively, can be defined if measures of $d$-variate association (see, e.g., \cite{Schmid2007}) are applied to $f_1(\mathbf{U}_1), f_2(\mathbf{U}_2),\dots,f_d(\mathbf{U}_d)$ for appropriate functions $f_1, \dots, f_d.$

\section{Acknowledgements}

Funding for Johan Segers' research was provided by IAP research network grant P6/03 of the Belgian government (Belgian Science Policy) and by ``Pro\-jet d'ac\-tions de re\-cher\-che con\-cer\-t\'ees'' number 07/12/002 of the Com\-mu\-nau\-t\'e fran\-\c{c}ai\-se de Bel\-gi\-que, granted by the Aca\-d\'e\-mie uni\-ver\-si\-taire de Lou\-vain. Funding in support of Julius Schnieders' work was provided by the Deutsche Forschungsgemeinschaft (DFG).
Morever, we are grateful to the Regional Computing Center at the University of Cologne for providing the computational resources required.

\section*{References}


\section{Appendix I}

\begin{enumerate}
\item Let $\pi _{p}\left( \mathbf{U}\right) = \prod_{i=1}^{p} \left( 1-U_{i}\right) $ \ and \ $\pi _{q}\left( \mathbf{V}\right) =%
\prod_{j=1}^{q}\left( 1-V_{j}\right) $

Then%
\begin{align*}
&\cov\left( \pi _{p}\left( \mathbf{U}\right) ,\pi _{q}\left( \mathbf{V}\right)
\right) =E_{C}\left( \pi _{p}\left( \mathbf{U}\right) ,\pi _{q}\left(
\mathbf{V}\right) \right) -E_{A}\left( \pi _{p}\left( \mathbf{U}\right)
\right) E_{B}\left( \pi _{q}\left( \mathbf{V}\right) \right) \\
&= \int_{[0,1]^p}\int_{[0,1]^q} \pi _{p}\left( \mathbf{u}\right) \pi _{q}\left( \mathbf{v}\right)dC(\mathbf{u},\mathbf{v}) - \int_{[0,1]^p}\pi _{p}\left( \mathbf{u}\right)dA(\mathbf{u}) \int_{[0,1]^q}\pi _{q}\left( \mathbf{v}\right)dB(\mathbf{v})
\\
&= \int_{[0,1]^p}\int_{[0,1]^q} C(\mathbf{u},\mathbf{v})d\mathbf{u}d\mathbf{v} - \int_{[0,1]^p}A(\mathbf{u})d\mathbf{u} \int_{[0,1]^q}B(\mathbf{v})d\mathbf{v} \\
&= \int_{[0,1]^p}\int_{[0,1]^q} \left( C(\mathbf{u},\mathbf{v})- A(\mathbf{u})B(\mathbf{v})          \right)d(\mathbf{u},\mathbf{v})
\end{align*}

\item Consider $A\left( \mathbf{U}\right) =F_{\mathbf{X}}\left( \mathbf{X}%
\right) $ \ and \ $B\left( \mathbf{V}\right) =F_{\mathbf{Y}}\left( \mathbf{Y}%
\right) .$

We then have%
\begin{equation*}
\cov\left( A\left( \mathbf{U}\right) ,B\left( \mathbf{V}\right) \right)
=E_{C}\left( A\left( \mathbf{U}\right) B\left( \mathbf{V}\right) \right)
-E_{A}\left( A\left( \mathbf{U}\right) \right) E_{B}\left( B\left( \mathbf{V}%
\right) \right)
\end{equation*}
and
\begin{eqnarray*}
E_{C}\left( A\left( \mathbf{U}\right) B\left( \mathbf{V}\right) \right)
& = & \int_{[0,1]^p}\int_{[0,1]^q}A\left( \mathbf{u}%
\right) B\left( \mathbf{v}\right) dC\left( \mathbf{u},\mathbf{v}\right) \\
& = & \int_{[0,1]^p}\int_{[0,1]^q}\overline{C}%
\left( \mathbf{u},\mathbf{v}\right) dA\left( \mathbf{u}\right) dB\left(
\mathbf{v}\right).
\end{eqnarray*}
Further
\begin{equation*}
E_{A}\left( A\left( \mathbf{U}\right) \right) =\int_{[0,1]^p}%
A\left( \mathbf{u}\right) dA\left( \mathbf{u}\right) =\int_{[0,1]^p} \overline{A}\left( \mathbf{u}\right) dA\left( \mathbf{u}\right)
\end{equation*}%
and%
\begin{equation*}
E_{B}\left( B\left( \mathbf{U}\right) \right) =\int_{[0,1]^q}%
B\left( \mathbf{u}\right) dB\left( \mathbf{u}\right) =\int_{[0,1]^q} \overline{B}\left( \mathbf{u}\right) dB\left( \mathbf{u}\right) .
\end{equation*}
\end{enumerate}

Therefore%
\begin{equation*}
\cov\left( A\left( \mathbf{U}\right) ,B\left( \mathbf{V}\right) \right) =%
\int_{[0,1]^p}\int_{[0,1]^q}\left( \overline{C}%
\left( \mathbf{u},\mathbf{v}\right) -\overline{A}\left( \mathbf{u}\right)
\overline{B}\left( \mathbf{v}\right) \right) dA\left( \mathbf{u}\right)
dB\left( \mathbf{v}\right) .
\end{equation*}%
The measure to be defined is therefore based on the weighted difference%
\begin{equation*}
\overline{C}\left( \mathbf{u},\mathbf{v}\right) -\overline{A}\left( \mathbf{u%
}\right) \overline{B}\left( \mathbf{v}\right) =P\left( \mathbf{U}\geq
\mathbf{u},\mathbf{V}\geq \mathbf{v}\right) -P\left( \mathbf{U}\geq \mathbf{u%
}\right) p\left( \mathbf{V}\geq \mathbf{v}\right)
\end{equation*}%
where the weights are given by $A$ and $B$.

Further
\begin{eqnarray*}
\var \left( A\left( \mathbf{U}\right) \right)
& = & E_{A}\left( A\left(
\mathbf{U}\right) A\left( \mathbf{U}\right) \right) -\left( E_{A}\left(
A\left( \mathbf{U}\right) \right) \right) ^{2} \\
& = & \int_{[0,1]^p}A\left( \mathbf{u}\right) A\left( \mathbf{u}%
\right) dA\left( \mathbf{u}\right) -\left( \int_{[0,1]^p}%
A\left( \mathbf{u}\right) dA\left( \mathbf{u}\right) \right) ^{2} \\
& = & \int_{[0,1]^p}\int_{[0,1]^p}A\left( \mathbf{u%
}\right) A\left( \mathbf{u}^{\prime }\right) dA\left( \mathbf{u\wedge u}%
^{\prime }\right) \\
&& \qquad \mbox{} -\int_{[0,1]^p}\int_{[0,1]^p}%
A\left( \mathbf{u}\right) A\left( \mathbf{u}^{\prime }\right) dA\left(
\mathbf{u}\right) dA\left( \mathbf{u}^{\prime }\right) \\
& = & \int_{[0,1]^p}\int_{[0,1]^p}\left( \overline{%
A}\left( \mathbf{u\vee u}^{\prime }\right) -\overline{A}\left( \mathbf{u}%
\right) \overline{A}\left( \mathbf{u}^{\prime }\right) \right) dA\left(
\mathbf{u}\right) dA\left( \mathbf{u}^{\prime }\right).
\end{eqnarray*}


\begin{thebibliography}{23}
\expandafter\ifx\csname natexlab\endcsname\relax\def\natexlab#1{#1}\fi
\expandafter\ifx\csname url\endcsname\relax
  \def\url#1{\texttt{#1}}\fi
\expandafter\ifx\csname urlprefix\endcsname\relax\def\urlprefix{URL }\fi

\bibitem[{Beran et~al.(2007)Beran, Bilodeau, and Lafaye~de
  Micheaux}]{Beran_etal_2007}
Beran, R., Bilodeau, M., Lafaye~de Micheaux, P., 2007. Nonparametric tests of
  independence between random vectors. Journal of Multivariate Analysis 98~(9),
  1805--1824.

\bibitem[{Cherubini et~al.(2004)Cherubini, Luciano, and
  Vecchiato}]{Cherubinietal2004}
Cherubini, U., Luciano, E., Vecchiato, W., 2004. Copula methods in finance.
  Wiley.

\bibitem[{Clayton(1978)}]{Clayton1978}
Clayton, D., 1978. A model for association in bivariate life tables and its
  application in epidemiological studies of familial tendency in chronic
  disease incidence. Biometrika 65, 141--151.

\bibitem[{Deheuvels(1979)}]{Deheuvels1979}
Deheuvels, P., 1979. La fonction de d\'{e}pendance empirique et ses
  propri\'{e}t\'{e}s. Acad. Roy. Belg. Bull. Cl. Sci. 65~(5), 274--292.

\bibitem[{Embrechts et~al.(2002)Embrechts, McNeil, and
  Straumann}]{Embrechts2002}
Embrechts, P., McNeil, A., Straumann, D., 2002. Risk Management: Value at Risk
  and Beyond. Cambridge University Press, Ch. Correlation and dependency in
  risk management: properties and pitfalls, pp. 176--223.

\bibitem[{Escoufier(1973)}]{Escoufier1973}
Escoufier, Y., 1973. Le traitment des variables vectorielles. Biometrics
  29~(4), 751--760.

\bibitem[{Fermanian et~al.(2004)Fermanian, Radulovi\'{c}, and
  Wegkamp}]{Fermanian2004}
Fermanian, J.-D., Radulovi\'{c}, D., Wegkamp, M., 2004. Weak convergence of
  empirical copula processes. Bernoulli 10~(5), 847--860.

\bibitem[{Hotelling(1936)}]{Hotelling1936}
Hotelling, H., 1936. Relations between two sets of variates. Biometrika 28,
  321--377.

\bibitem[{Joe(1997)}]{Joe1997}
Joe, H., 1997. Multivariate Models and Dependence Concepts. Chapman \& Hall,
  London.

\bibitem[{Kojadinovic and Holmes(2009)}]{Kojadinovic2009}
Kojadinovic, I., Holmes, M., 2009. Tests of independence among continuous
  random vectors based on cram\'{e}r-von mises functionals of the empirical
  copula process. Journal of Multivariate Analysis 100, 1137--1154.

\bibitem[{Nelsen(2006)}]{Nelsen2006}
Nelsen, R.~B., 2006. An Introduction to Copulas, 2nd Edition. Springer Series
  in Statistics. Springer, New York.

\bibitem[{Nelsen et~al.(2003)Nelsen, Quesada-Molina, Rodríguez-Lallena, and
  Úbeda Flores}]{Nelsen2003}
Nelsen, R.~B., Quesada-Molina, J.~J., Rodríguez-Lallena, J.~A., Úbeda Flores,
  M., 2003. Kendall distribution functions. Statistics \& Probability Letters
  65~(3), 263 -- 268.

\bibitem[{Quessy(2010)}]{Quessy2010}
Quessy, J., 2010. Applications and asymptotic power of marginal-free tests of
  stochastic vectorial independence. Journal of Statistical Planning and
  Inference 140, 3058--3075.

\bibitem[{R\'{e}millard(2009)}]{Remillard2009b}
R\'{e}millard, B., 2009. Discussion of: Brownian distance covariance. The
  Annals of Applied Statistics 3~(4), 1295--1298.

\bibitem[{Robert and Escoufier(1976)}]{Robert1976}
Robert, P., Escoufier, Y., 1976. A unifying tool for linear multivariate
  statistical methods: the rv-coefficient. Appl. Statist. 25~(3), 257--265.

\bibitem[{R{\"{u}}schendorf(1976)}]{rueschendorf76}
R{\"{u}}schendorf, L., 1976. Asymptotic distributions of multivariate rank
  order statistics. Annals of Statistics 4~(5), 912--923.

\bibitem[{Schmid et~al.(2009)Schmid, Schmid, Blumentritt, Gaisser, and
  Ruppert}]{Schmid2009}
Schmid, F., Schmid, R., Blumentritt, T., Gaisser, S., Ruppert, M., 2009.
  Copula-based measures of multivariate association. In: Jaworski, P., Durante,
  F., Härdle, W., Rychlik, T. (Eds.), Copula Theory and its applications.
  Lecture Notes in Statistics - Proceedings.

\bibitem[{Schmid and Schmidt(2007)}]{Schmid2007}
Schmid, F., Schmidt, R., 2007. Multivariate extensions of spearman's rho and
  related statistics. Statistics Probability Letters 77~(4), 407--416.

\bibitem[{Segers(accepted)}]{segers10}
Segers, J., accepted. Asymptotics of empirical copula processes under
  nonrestrictive smoothness assumptions. Bernoulli, ArXiv:1012.2133v2.

\bibitem[{Sklar(1959)}]{Sklar1959}
Sklar, A., 1959. Fonctions de r\'{e}paration \`{a} n dimensions et leurs
  marges. Publ. Inst. Statist. Univ. Paris 8, 229--231.

\bibitem[{Sz\'{e}kely and Rizzo(2009)}]{Szekely2009}
Sz\'{e}kely, G., Rizzo, M., 2009. Brownian distance covariance. The Annals of
  Applied Statistics 3~(4), 1236--1265.

\bibitem[{Sz\'{e}kely et~al.(2007)Sz\'{e}kely, Rizzo, and
  Bakirov}]{Szekely2007}
Sz\'{e}kely, G., Rizzo, M., Bakirov, N., 2007. Measuring and testing dependence
  by correlation of distances. Ann. Statist. 35~(6), 2769--2794.

\bibitem[{Van~der Vaart and Wellner(1996)}]{vaartwellner96}
Van~der Vaart, A.~W., Wellner, J.~A., 1996. Weak convergence and empirical
  processes. Springer Verlag, New York.

\end{thebibliography}
\end{document}